\documentstyle[12pt,bezier]{article}
\newdimen\lineskp
\newcommand{\mabel}[1]{\label{#1}}
\newcommand{\mibitem}[1]{\bibitem{#1}}

\begin{document}
\baselineskip= \lineskp plus 1pt minus 1pt

\begin{titlepage}

\rightline{hep-th/9605172} 
\rightline{May 22, 1996}

\vskip 0.1truein
\begin{center}

{\bf Exact Solution of the One-Dimensional Non-Abelian Coulomb Gas at
Large $N$}\\
\vskip 0.3truein
{\bf G.W. Semenoff, O. Tirkkonen}\\
\medskip
{\it Department of Physics and Astronomy,
University of British Columbia\\
Vancouver, British Columbia, Canada V6T 1Z1}\\
\mbox{} \\ { and} \\
\mbox{} \\
{\bf K. Zarembo}\\
 \mbox{} \\ {\it Steklov Mathematical Institute,} \\
{\it Vavilov st. 42, GSP-1, 117966 Moscow, RF}
\\and  \\ {\it Institute of Theoretical and Experimental Physics,}
\\ {\it B. Cheremushkinskaya 25, 117259 Moscow, RF} \\
\end{center}
\medskip
\begin{center}

{\bf Abstract}
\end{center}

\noindent
The problem of computing the thermodynamic properties of a
one-dimensional gas of particles which transform in the adjoint
representation of the gauge group and interact through non-Abelian
electric fields is formulated and solved in the large $N$ limit.  The
explicit solution exhibits a first order confinement-deconfinement
phase transition with computable properties and describes two
dimensional adjoint QCD in the limit where matter field masses are
large.

\vfill

\centerline{PACS: 11.10.Wx; 11.15.-q}

\end{titlepage}

\noindent
Two dimensional quantum chromodynamics (QCD) with adjoint
representation matter fields is the simplest field theoretical model
which exhibits some of the important common features of string theory
and the confining phases of gauge theory. Most notable are the
infinite number of asymptotically linear Regge trajectories and a
density of states which increases exponentially with energy.  In two
dimensions, the Yang-Mills field itself has no propagating degrees of
freedom.  In adjoint QCD, the matter fields provide dynamics by
playing a role analogous to the transverse gluons of higher
dimensional gauge theory.  In fact,  dimensional
reduction of three dimensional Yang Mills theory produces two
dimensional QCD with massless adjoint scalar quarks.
Moreover, since adjoint matter fields do not decouple in the infinite $N$
limit, the large $N$ expansion is of a similar level of complexity to
that of higher dimensional Yang-Mills theory.  One would expect it
to exhibit some of the stringy features of the confining phase which
are emphasized in that limit.

Although adjoint QCD is not explicitly
solvable, even at infinite $N$, details of its spectrum were readily
analyzed by approximate and numercial techniques
~\cite{dk,kutasov,Bhanot,kogan}.  In addition, Kutasov ~\cite{kutasov}
exploited an argument which was originially due to Polchinski
~\cite{polchinski} to show that the confining phase must be unstable
at high temperature and suggested it as a tractable model where the
confinement-deconfinement transition could be investigated.

In this letter, we shall formulate and find an explicit solution of
the large N limit of a simplfied version of adjoint QCD.  We shall
consider a one-dimensional gas of non-dynamical particles which have
adjoint color charges and which interact with each other through
non-Abelian electric fields.  Because there are no dynamical gluons
which could screen adjoint charges in one dimension, at low
temperature and density, adjoint quarks are confined\footnote{In
higher dimensions, an adjoint charge and a gluon could form a color
singlet bound state.}. They form colorless ``hadron'' bound states
with two or more adjoint quarks connected by non-dynamical strings of
electric flux.  The large $N$ limit resembles a non-interacting string
theory in that, at infinite $N$, the energy of a state is proportional
to the total length of all strings of electric flux plus a chemical
potential times the total number of quarks.  The property of
confinement is defined by estimating the energy required to introduce
an external fundamental representation quark-antiquark pair into the
system.  In the confining phase, where the hadron gas is dilute, the
quark-antiquark energy is proportional to the length of the electric
flux string which, to obtain gauge invariance, must connect them.
This gives them a confining interaction.  Some typical confined
configurations are depicted in Fig.  1.  In the confined phase, the
average particle number density and the energy density are small --- in
the large $N$ limit both are of order one, rather than $N^2$ which one
would expect from naive counting of the degrees of freedom.  This is
consistent with the fact that in a confining phase the number of
degrees of freedom, i.e. hadrons, is independent of $N$.  In contrast,
in the deconfined phase, since the number of degrees of freedom,
i.e. quarks and gluons, is proportional to $N^2$ the particle density
and energy are also of order $N^2$.

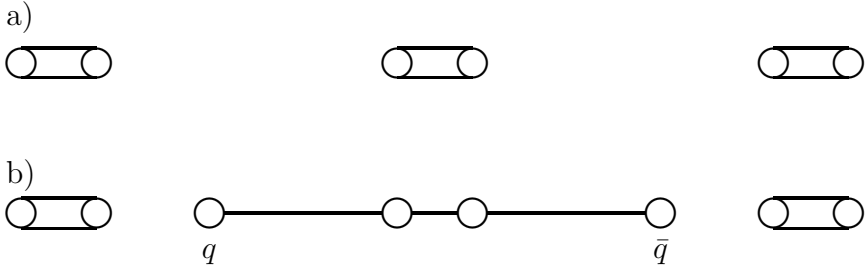
\begin{figure}[tb]
\setlength{\unitlength}{1mm}
\begin{picture}(118,37)
\thicklines

\put(3,33){a)}

\put(5,28){\circle{4}}
\put(5,30){\line(1,0){10}}
\put(5,26){\line(1,0){10}}
\put(15,28){\circle{4}}

\put(55,28){\circle{4}}
\put(55,30){\line(1,0){10}}
\put(55,26){\line(1,0){10}}
\put(65,28){\circle{4}}

\put(105,28){\circle{4}}
\put(105,30){\line(1,0){10}}
\put(105,26){\line(1,0){10}}
\put(115,28){\circle{4}}


\put(3,12){b)}

\put(5,8){\circle{4}}
\put(5,10){\line(1,0){10}}
\put(5,6){\line(1,0){10}}
\put(15,8){\circle{4}}

\put(30,8){\circle{4}}
\put(29,2){$q$}
\put(32,8){\line(1,0){21}}
\put(55,8){\circle{4}}
\put(57,8){\line(1,0){6}}
\put(65,8){\circle{4}}
\put(67,8){\line(1,0){21}}
\put(90,8){\circle{4}}
\put(89,2){$\bar q$}

\put(105,8){\circle{4}}
\put(105,10){\line(1,0){10}}
\put(105,6){\line(1,0){10}}
\put(115,8){\circle{4}}

\end{picture}
 \caption{ {\it Some examples of states in the confining phase.  Each
adjoint quark connects with two electric flux strings and fundamental
quarks in (b) connect with one string.  }}

\smallskip

\end{figure}

As temperature or density is increased, eventually we arrive at the
situation where there are electric flux strings almost everywhere.
Then, adding an additional flux string, or modifying the existing
network of strings to accommodate a fundamental representation
quark-antiquark pair, involves a negligibly small addition to the
energy of the total system (see Fig. 2).  This is typical of the
deconfined phase.

\begin{figure}[tb]
\setlength{\unitlength}{1mm}
\begin{picture}(124,37)
\thicklines

\put(0,33){a)}

\put(2,28){\circle{4}}
\put(2,30){\line(1,0){10}}
\put(2,26){\line(1,0){10}}
\put(12,28){\circle{4}}

\put(27,28){\circle{4}}
\put(29,28){\line(1,0){6}}
\put(37,28){\circle{4}}
\put(39,28){\line(1,0){6}}
\put(47,28){\circle{4}}
\put(49,28){\line(1,0){6}}
\put(57,28){\circle{4}}
\bezier{100}(28.5,29.5)(42,37)(55.5,29.5)

\put(67,28){\circle{4}}
\put(69,28){\line(1,0){6}}
\put(77,28){\circle{4}}
\put(79,28){\line(1,0){6}}
\put(87,28){\circle{4}}
\bezier{80}(68.5,29.5)(77,35)(85.5,29.5)

\put(102,28){\circle{4}}
\put(104,28){\line(1,0){6}}
\put(112,28){\circle{4}}
\put(114,28){\line(1,0){6}}
\put(122,28){\circle{4}}
\bezier{80}(103.5,29.5)(112,35)(120.5,29.5)


\put(0,13){b)}

\put(2,8){\circle{4}}
\put(2,10){\line(1,0){10}}
\put(2,6){\line(1,0){10}}
\put(12,8){\circle{4}}

\put(21,8){\circle{4}}
\put(20,2){$q$}
\put(23,8){\line(1,0){5}}
\put(30,8){\circle{4}}
\put(32,8){\line(1,0){5}}
\put(39,8){\circle{4}}
\put(41,8){\line(1,0){5}}
\put(48,8){\circle{4}}
\put(50,8){\line(1,0){5}}
\put(57,8){\circle{4}}
\put(59,8){\line(1,0){5}}
\put(66,8){\circle{4}}
\put(68,8){\line(1,0){5}}
\put(75,8){\circle{4}}
\put(77,8){\line(1,0){5}}
\put(84,8){\circle{4}}
\put(86,8){\line(1,0){5}}
\put(93,8){\circle{4}}
\put(92,2){$\bar q$}

\put(102,8){\circle{4}}
\put(104,8){\line(1,0){6}}
\put(112,8){\circle{4}}
\put(114,8){\line(1,0){6}}
\put(122,8){\circle{4}}
\bezier{80}(103.5,9.5)(112,15)(120.5,9.5)

\end{picture}
\caption{ {\it Typical configurations  in the
deconfined phase.  }}

\smallskip

\end{figure}
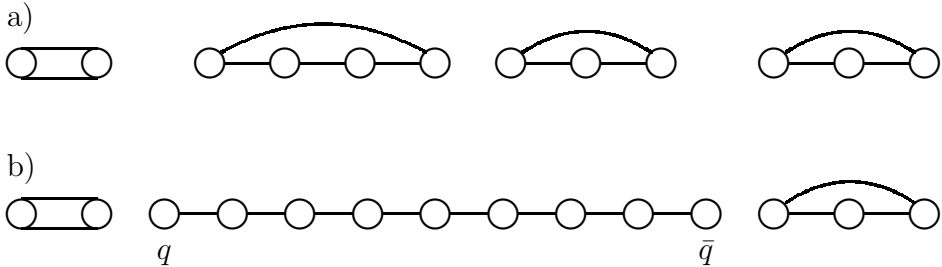

Between these two phases is a transition, which we shall show in this
paper, is of first order.  In the string picture, this phase
transition occurs when the strings in a typical configuration
percolate in the one-dimensional space.  The order parameter is the
Polyakov loop operator ~\cite{pol,sus} which measures the exponential
of the negative of the free energy which is required to insert a
single, unpaired fundamental representation quark source into the
system.  This free energy is infinite (and the expectation value of
the Polyakov loop is zero) in the confining phase and it is finite in
the deconfined phase.

In the Hamiltonian formulation of two dimensional Yang-Mills theory,
the electric field is the canonical conjugate of the spatial component
of the gauge field, $\left[ A^a(x), E^b(y) \right]~=~i\delta^{ab}(x-y)$.
The Hamiltonian is\footnote{Here,
for concreteness, we consider $U(N)$ gauge theory. The gauge field $A
= A^a t_a$, with $t_a$ the generators in the fundamental representation. }
\begin{equation}
H=\int dx~ \frac{e^2}{2} \sum_{a=1}^{N^2} (E^a(x))^2 \ ,
\mabel{ham}
\end{equation}
and the Gauss' law constraint, which takes the form of a physical
state condition, is
\begin{equation}
\left( \frac{d}{dx}E^a(x) -
f^{abc}A^b(x)E^c(x) + \sum_{i=1}^K T^a_i\delta(x-x_i)\right)~\Psi_{\rm
phys.}~=~ 0 \ .
\mabel{gauss}
\end{equation}
There are particles with adjoint representation color charges located
at positions $x_1,\ldots ,x_K$.  $T^a_i$ are generators in the adjoint
representation operating on the color degrees of freedom of the {\it i}'th
particle.

In the functional Schr\"odinger picture, the states are functionals of
the gauge field and the electric field is the functional derivative
operator $ E^a(x)~=~\frac{1}{i}\frac{\delta}{\delta A^a(x)} $.  The
functional Schr\"odinger equation is that of a free particle
\begin{equation}
\int dx\left( -\frac{e^2}{2}\sum_{a=1}^{N^2}
\frac{\delta^2}{(\delta A^a(x))^2}\right)
~\Psi^{a_1\ldots a_K}\left[A;x_1, \ldots, x_K\right]~
  =~{\cal E}~\Psi^{a_1\ldots
a_K}\left[A;x_1, \ldots, x_K\right]
\end{equation}
Gauss' law implies that the physical states, i.e. those which obey the
gauge constraint (\ref{gauss}), transform as
\begin{equation}
\Psi^{a_1\ldots a_K}\left[ A^g;x_1, \ldots, x_K\right]=
g^{\rm Ad}_{a_1b_1}(x_1)\ldots g^{\rm Ad}_{a_Kb_K}(x_K)
\Psi^{b_1\ldots b_K}\left[A;x_1, \ldots, x_K\right]
\end{equation}
where $A^g\equiv gAg^{\dagger}+ig\nabla g^{\dagger}$ is the gauge
transform of $A$.  For a fixed number of external charges, this model
is explicitly solvable.  In the following we shall examine its
thermodynamic featurs, where we assume that the particles have
Maxwell-Boltzmann statistics.

We find it convenient to work with the grand canonical ensemble.  The
partition function is constructed by taking the trace of
the Gibbs density $e^{-H/T}$ over physical states.  This can be
implemented by considering eigenstates of $A^a(x)$ (and
an appropriate basis for the non-dynamical particles) $\vert A \rangle
e_{a_1}\ldots e_{a_K}$.  Projection onto gauge invariant states
involves a projection operator which has the net effect of gauge
transforming the state field at one side of the trace, and integrating
over all gauge transformations \cite{gpy}. The resulting partition
function is
\begin{equation}
Z[x_i,T]=\int[dA][dg]~\left<A\right| e^{-H/T} \left|A^g\right>
{\rm Tr}~g^{\rm Ad}(x_1)\ldots
{\rm Tr}~g^{\rm Ad}(x_K) \ ,
\end{equation}
where $[dg(x)]$ is the Haar measure on the space of mappings from the
line to the group manifold and $[dA]$ is a measure on the convex
Euclidean space of gauge field configurations.  For U(N), the trace in
the adjoint representation is ${\rm Tr}~g^{\rm Ad}(x) = \left| {\rm
Tr}~ g(x)\right|^2$ where $g(x)$ is in the fundamental
representation. In order to form the grand canonical ensemble, we
average over the particle positions by integrating $\int dx_1\ldots
\int dx_K$, multiply by the fugacity to the power $K$, $\lambda^K$,
divide by the statistics factor $1/K!$ and sum over $K$.  The result
is
\begin{equation}
Z[\lambda,T]~=~\int [dA][dg]~e^{-S_{\rm eff}[A,g]}
\mabel{partition}
\end{equation}
where the effective action is
\begin{equation}
e^{-S_{\rm eff}[A,g]}~=~ \left< A\right| e^{-H/T}\left| A^g
\right>~\exp\left(\int dx~\lambda\left|{\rm Tr}~ g\right|^2 \right)
\end{equation}
The Hamiltonian is the Laplacian on the space of gauge fields.  Using
the explicite form of the  heat kernel
$$\left< A\right| e^{-H/T} \left| A^g\right> \sim
\exp\left(-\int dx ~\frac{T}{e^2}~{\rm Tr}~(A-A^g)^2\right)\ ,$$
we see that the effective theory is the gauged principal chiral model
with a quadratic potential
\begin{equation}
S_{\rm eff}[A,g]~=~ \int dx~\left( \frac{T}{e^2}~ {\rm Tr}\left|
\nabla g + i[A,g]\right|^2 -\lambda\left| {\rm
Tr}~g\right|^2
 \right) \mabel{seff}
\end{equation}
This effective action with $\lambda=0$ was discussed by Grignani et.al.
~\cite{gsst}.  It is gauge invariant, $S_{\rm eff}[A,g]~=~ S_{\rm
eff}[A^h, hgh^{\dagger}] $,
and has the global symmetry
$S_{\rm eff}[A,g]~=~ S_{\rm eff}[A,~z~g]$,
where $z$ is a constant element from the center of the gauge group,
which for U(N) is U(1) and for  SU(N) is $Z_N$.

The realization of this center symmetry governs confinement
~\cite{pol,sus}.  When the symmetry is represented faithfully, the
theory is in the confining phase.  The Polyakov loop operator ${\rm
Tr}~ g(x)$ transforms under the center as ${\rm Tr}~g(x)~\rightarrow
z~{\rm Tr}~g(x) $. Thus the expectation value of the Polyakov loop
operator must average to zero if the symmetry is not spontaneously
broken.  This expectation value is interpreted as the free energy of
the system with an additional external charge in the fundamental
representation of the gauge group situated at point $x$,
$F[x,\lambda,T]=-T\ln\left< {\rm Tr}~g(x)\right>$.  For finite $N$,
and $D=3$, ideas of universality have been applied to study phase
transitions with this order parameter in SU(N) gauge theory
~\cite{sy}.  The phase transition should be second order for $N=2$ and
first order for $N>2$.

Here we amalyze the effective theory (\ref{partition}),(\ref{seff}) in
the large $N$ limit.  If we rescale the coupling constant so that
$\frac{e^2}{T}\equiv \frac{2 \gamma}{N}$, both terms in the action
(\ref{seff}) are of order $N^2$ and in the large $N$ limit the
partition function is dominated by the configuration which minimizes
the action.  Gauge invariance can be used to diagonalize the matrices
$g_{ij}(x)= e^{i\alpha_i(x)}\delta_{ij}$. The density of eigenvalues
$\rho(\theta,x)~=~\frac{1}{N}\sum_{i=1}^N \delta(\theta-\alpha_i(x)) $
corresponding to the large $N$ saddlepoint now characterizes the
properties of the system.  A constant density $ \rho_{\rm
conf.}(\theta,x)~=~ \frac{1}{2\pi}$ realizes the center symmetry, and
thus corresponds to the confining phase. A density peaked at some
value of $\theta$ explicitely breaks the center symmetry, and
corresponds to a deconfined phase.

If, in the general case, we consider the Fourier expansion
\begin{equation}
\rho(\theta,x)~=~\frac{1}{2\pi}+\frac{1}{2\pi}
\sum_{n\neq 0}c_n(x) e^{-in\theta}
~~~,~~c_n(x)^*= c_{-n}(x) ~,
\mabel{den}
\end{equation}
the Fourier coefficients $c_n(x)$ characterize the possible deconfined
phases of the theory.  If one of them were non-zero, we would have in
the infinite $N$ limit $\frac{1}{N}\left< {\rm Tr}~g^n(x) \right> ~=~
c_n(x) $.  This would indicate that a composite of $n$ fundamental
quarks would have finite free energy and would not be confined.

In order to find the configurations of the eigenvalue density
(\ref{den}) that minimize the action, we shall use the collective
field theory approach of Refs.  \cite{JS,Wadia,Zar}.  Alternatively to
the gauge fixing that we have discussed, we consider (\ref{partition})
in the gauge $A=0$ which can be fixed on the open line.  Then the
thermodynamic problem is equivalent to unitary matrix quantum
mechanics
\begin{equation}
Z[\lambda,T]~=~\int [dg]~\exp\left(-\int dx\left(\frac{N}{2 \gamma}{\rm
Tr}\left| \nabla g
\right|^2-\lambda\left|{\rm Tr}~g\right|^2 \right)\right)
\end{equation}
This model can be solved in the large $N$ limit by the methods of
collective field theory.  The method is essentially based on the
relation between matrix quantum mechanics and nonrelativistic
fermions \cite{BIPZ}.  In the large $N$ limit the eigenvalue density obeys a
classical, saddle point equation which can be deduced from canonical
analysis of the collective field theory Hamiltonian \cite{JS}
\begin{equation}
H~=~\int d\theta\left[ \frac{\gamma}{2}\rho(\theta)\left(
\frac{\partial \pi}{\partial \theta}\right)^2 
 +\frac{\pi^2\gamma}{6}\rho^3(\theta) \right]
 - \lambda\left| \int d\theta~\rho(\theta)e^{i\theta}\right|^2
 - \frac{\gamma}{24} \ , \mabel{col}
\end{equation}
 and subsequent Wick rotation to imaginary time.  Here $\Pi(\theta)$
is the variable which is the canonical conjugate of $\rho(\theta)$, so
that the Poisson bracket is $ \left\{ \rho(\theta), \Pi(\theta')
\right\}~=~ \delta(\theta - \theta') $.  The velocity of the Fermi
fluid is $v(\theta)=\partial\Pi/\partial\theta$.  In the equations of
motion following from (\ref{col}), we change $t\rightarrow ix$,
$v\rightarrow -iv$ and obtain
\begin{equation}
\frac{\partial\rho}{\partial x}+\gamma\frac{\partial}{\partial\theta}
(\rho v)=0
\mabel{coll1}
\end{equation}
\begin{equation}
\frac{\partial v}{\partial x}+\gamma v\frac{\partial v}{\partial\theta}
-\pi^2\gamma\rho \frac{\partial\rho}{\partial\theta}+ 2\lambda{\rm Im}
\left(e^{-i\theta}c_1(x)\right)~=~0 \ .
\mabel{coll2}
\end{equation}
 It is expected that the solution of these
equations corresponding to an equilibrium state of the system is a
constant $\rho(x,\theta) = \rho_0(\theta)$.  At least at sufficiently
low temperature or, equivalently, at sufficiently large $\gamma$, the
system is in the confining phase with unbroken center symmetry, so
that $\rho_0=\rho_{\rm conf.}=1/2\pi$.  This is always a solution of
the equations of motion (\ref{coll1}),(\ref{coll2}) since $c_1=0$.

However, this solution is stable against small fluctuations only if
$\gamma$ is large enough.  To find the spectrum of excitations in this
phase, we linearize the equations of motion around $\rho_{{\rm
conf.}}$.  To do this, we consider the $c_n(x)$ of eq. (\ref{den})
and  $v$ infinitesimal.  The resulting equation for $c_n(x)$ is
\begin{equation}
\left(-\nabla^2+
\frac{\gamma^2 n^2}{4}-\lambda\gamma(\delta_{n,1}
+ \delta_{n,-1})\right)~c_n(x)
=0~~;~~n\neq 0 \ .
\end{equation}
At $\gamma=\gamma_c(\lambda)=4\lambda $,
the lowest eigenvalue corresponding to $n=\pm 1$ goes to zero.  For
smaller $\gamma$ this eigenvalue is negative and the strong coupling
solution is unstable with $c_{\pm 1}$ the first modes to become unstable.
However, for reasons which will become clear once we consider the weak
coupling phase, $\gamma_c(\lambda)$ should not be identified with the
point of the deconfining phase transition.

The solution in the deconfined phase can be obtained by integration of
eq.  (\ref{coll2}) at $v=0$.  The density $\rho_0(\theta)$ can always
be chosen to be an even function of $\theta$.  Thus $c_1$ is real, and
one finds from eq. (\ref{coll1}): 
$$\rho_0(\theta)=\frac{1}{\pi}\sqrt{
\frac{2}{\gamma}E+2\lambda c_1 \cos\theta} \ .$$
Outside the region $[-\theta_{\rm max}, \theta_{\rm max}]$ with
$\theta_{\rm max}=\pi-\arccos\frac{E}{2\lambda c_1} $, the density
$\rho(\theta)$ is zero.

The Fermi energy $E$ and the constant $c_1$ are to be determined from
the normalization condition $\int_{-\theta_{\rm max}}^{~\theta_{\rm
max}}d\theta ~\rho_0(\theta;E,c_1) = 1$, and the consistency condition
$\int_{-\theta_{\rm max}}^{~\theta_{\rm max}}d\theta\cos\theta
~\rho_0(\theta;E,c_1) = c_1$ derived from eq. (\ref{den}).  It follows
from these equations that $\theta_{\rm max}$ tends to zero at $\gamma
\rightarrow 0$ and grows with the increase of $\gamma$.  Eventually it
reaches $\pi$, where the weak coupling phase terminates, because the
eigenvalue distribution begins to overlap with itself due to
$2\pi$-periodicity.  At the critical point $E_*=2\lambda c_{1*}$, the
normalization and consistency integrals can be done explicitly. We
find that $c_{1*}=1/3$ and $ \gamma_*(\lambda)=\frac{128}{3\pi^2}
~\lambda \approx 4.324 ~\lambda$.

We obtain the following picture of the deconfining phase transition
(Fig.3).  The weak and strong coupling phases can coexist, because
$\gamma_c(\lambda) <\gamma_*(\lambda)$, although the region, where
both phases are stable is very narrow, since $\gamma_c(\lambda)$ and
$\gamma_*(\lambda)$ are numerically close to each other.  The phase
transition is of the first order and takes place at some
$\gamma_0(\lambda)$ between $\gamma_c(\lambda)$ and
$\gamma_*(\lambda)$.  At the point of the phase transition the
free energies of both phases are equal to each other.  Substituting
$\rho_0(\theta)$ into equation (\ref{col}) one can find the free
energy per unit volume, to leading order in the large $N$ limit,
\begin{equation}
\frac{F}{N^2}~=~\left\{ \matrix{ 0,& {\rm in~the~confining~phase}\cr
\frac{1}{3}E-\frac{1}{3}\lambda c_1^2 - \frac{\gamma}{24},~~~~~~~&
{\rm ~~in~the ~deconfined~phase}\cr}\right.
\end{equation}
The equations determining the critical line can be solved numerically
to obtain $ \gamma_0(\lambda)=4.219 ~\lambda$.

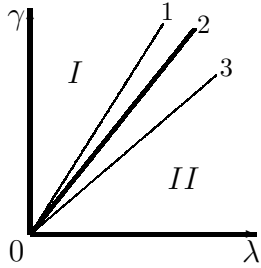
\begin{figure}[tbp]
\unitlength 0.3mm
\linethickness{0.4pt}
\begin{picture}(120.00,120.00)(-200,0)
\linethickness{1.6pt}
\put(120.00,20.00){\vector(1,0){0.2}}
\put(20.00,20.00){\line(1,0){100.00}}
\put(20.00,120.00){\vector(0,1){0.2}}
\put(20.00,20.00){\line(0,1){100.00}}
\linethickness{0.4pt}
\multiput(20.00,20.00)(0.12,0.19){492}{\line(0,1){0.19}}
\multiput(20.00,20.00)(0.14,0.12){592}{\line(1,0){0.14}}
\linethickness{1.6pt}
\multiput(20.00,20.00)(0.12,0.15){609}{\line(0,1){0.15}}
\linethickness{0.4pt}
\put(14.00,115.00){\makebox(0,0)[cc]{$\gamma$}}
\put(118.00,13.00){\makebox(0,0)[cc]{$\lambda$}}
\put(14.00,13.00){\makebox(0,0)[cc]{$0$}}
\put(40.00,92.00){\makebox(0,0)[cc]{$I$}}
\put(88.00,45.00){\makebox(0,0)[cc]{$II$}}
\put(81.00,119.00){\makebox(0,0)[cc]{{\footnotesize 1}}}
\put(97.00,113.00){\makebox(0,0)[cc]{{\footnotesize 2}}}
\put(107.00,94.00){\makebox(0,0)[cc]{{\footnotesize 3}}}
\end{picture}
\caption { {\it
 The large $N$ phase diagram of the one-dimensional model. $I$ --
strong coupling (confining) phase, $II$ --- weak coupling (deconfining)
phase; 1 --- line on which the weak coupling phase terminates:
$\gamma=\gamma_*(\lambda)$, 2 --- line of the first-order phase
transition: $\gamma =\gamma _0(\lambda )$, 3 --- line of the
instability of the strong coupling phase: $\gamma=\gamma_c(\lambda)$}
}
\smallskip
\end{figure}

The model which we have considered in this Section is adjoint QCD in
the limit where the particles are heavy.  The fugacity parameter can
be computed from a one-loop diagram as $ \lambda=
\sqrt{\frac{mT}{2\pi} }e^{-m/T}$, the exponential being simply the
Boltzmann weight of a particle with mass $m$.  It is assumed that $m>>T$
for classical statistical mechancis to be applicable and $m>>e$ to
suppress pair production. Our results indicate that the phase
transition is of first order with critical line approximately given by
the equation $\frac{e^2 N}{2T}~\approx~4.2\sqrt{ \frac{mT}{2\pi}
}e^{-m/T}$.  There exists a region of parameters in which this
equation has a solution and the conditions of applicability of our
simplified model are satisfied.

\noindent
The work of K. Zarembo was supported in part by INTAS grant 94-0840.
The work of G. Semenoff and O. Tirkkonen was supported in part by NSERC
of Canada.

\end{document}